\documentclass[11pt]{article}
\usepackage[includeheadfoot,margin=1in]{geometry}
\usepackage{times}
\usepackage{cite}
\usepackage{amsmath,amsfonts,amssymb, booktabs,url}
\usepackage{mathtools}
\usepackage{enumerate}
\usepackage{verbatim}
\usepackage{commath}
\usepackage{graphicx}
\usepackage{booktabs}
\usepackage{tabularx}
    \usepackage[usenames]{color}
    \definecolor{plum}  {rgb}{.4,0,.4}
    \definecolor{BrickRed} {rgb}{0.6,0,0}
	\definecolor{DarkBlue} {rgb}{0,0,0.6}
\usepackage[plainpages=false,pdfpagelabels,colorlinks=true,linkcolor=BrickRed,citecolor=plum,urlcolor=DarkBlue]{hyperref}
\usepackage[mathcal]{euscript}
\usepackage{cleveref}
\begin{document}
\title{\LARGE Resource-Aware Task Allocator Design: Insights and Recommendations for Distributed Satellite Constellations}
\author{\Large Bharadwaj Veeravalli \\ \href{mailto:elebv@nus.edu.sg}{elebv@nus.edu.sg} \\ %
Department of Electrical and Computer Engineering,\\ National University of Singapore, 4 Engineering Drive 3, Singapore.}
\date{\today}
\maketitle
\begin{abstract} We present the design of a Resource-Aware Task Allocator (RATA) and an empirical analysis in handling real-time tasks for processing on Distributed Satellite Systems (DSS). We consider task processing performance across low Earth orbit (LEO) to Low-Medium Earth Orbit (Low-MEO) constellation sizes, under varying traffic loads. Using Single-Level Tree Network(SLTN)-based cooperative task allocation architecture, we attempt to evaluate some key performance metrics - blocking probabilities, response times, energy consumption, and resource utilization across several tens of thousands of tasks per experiment. Our resource-conscious RATA monitors key parameters such as arrival rate, resources (on-board compute, storage, bandwidth, battery) availability, satellite eclipses' influence in processing and communications. This study is an important step towards analyzing the performance under lighter to stress inducing levels of compute intense workloads to test the ultimate performance limits under the combined influence of the above-mentioned factors. Results show pronounced non-linear scaling: while capacity increases with constellation size, blocking and delay grow rapidly, whereas energy remains resilient under solar-aware scheduling. The analysis identifies a practical satellite-count limit for baseline SLTNs and demonstrates that CPU availability, rather than energy, is the primary cause of blocking. These findings provide quantitative guidance by identifying thresholds at which system performance shifts from graceful degradation to collapse. 
\end{abstract}
\section{INTRODUCTION}
Low Earth orbit(LEO) satellite constellations are increasingly deployed for Earth observation, communications, and edge computing applications. Modern LEO satellites at 500-1,200 km altitude carry substantial computational resources (10-50 GFLOPS, 4-8 cores, 64-256 GB memory), enabling on-orbit processing that reduces downlink bandwidth requirements and provides low-latency services \cite{Giuff2020,DlPort2019}. 

In such satellite based compute platforms, resource optimization is a critical issue as there are several influencing parameters that affect the performance. In such resource scarce platforms, satisfying the needs in task processing is an extremely challenging problem as task types differ in their compute, storage, and energy  demands. For instance, the range of task sizes that arrive in real-time to distributed satellite systems(DSS) is between 2-15 GB and their compute intensities vary over a large range of 25M-1.25B FLOP/MB (demanding 5 to 40s processing on the lower side) to 25M-125M FLOP/MB (demanding 10-60s processing on higher side). Further this has a direct impact on energy consumption and satellites' battery recharging opportunities during their orbital periods. Thus, a comprehensive design of a resource aware task allocator taking into account of all the above-mentioned issues in DSS is a challenging problem. 

In the following section, we review the literature most relevant to our study, highlight remaining gaps, and show that effective task allocation in DSS must jointly optimize computation, communication, energy, and reliability, rather than treating these factors separately in real-world implementations. Significant results from this survey offer actionable guidance for constellation design based on comprehensive empirical evidence. Also, constellations scaling from tens to hundreds of satellites introduces complex resource management challenges: {\it How should tasks be allocated across distributed satellites with limited energy, storage, communication bandwidth, and computational capacity?} Thus, in this study, we address a fundamental question - {\it how does constellation size affect task processing performance under realistic LEO to Mid-MEO operational constraints?} We systematically evaluate processing tasks ranging from 2-15 GB under identical hardware specifications on four constellation scales (in LEO to Low-MEO tiers), revealing non-intuitive scaling laws and establishing an empirical as well as quantitative performance bounds for current architectures.
\subsection{Relevant Literature}
Earlier works by Giuffrida et al.\ (2020) demonstrated the feasibility of
deep neural network-based on-board processing using SmallSats equipped with
commercial-off-the-shelf processors, showing that on-orbit hyperspectral
image classification can selectively filter data before downlink, reducing the
volume of imagery transmitted to ground stations \cite{Giuff2020}. This work
helped shift system design from ground-centric processing toward hybrid
space--ground architectures where satellites perform meaningful computation
before relaying results. Extending this direction, Del Portillo et al.\ (2019)
conducted a technical comparison of three LEO mega-constellation systems for
global broadband and showed that computational capacity must scale
super-linearly with constellation size to preserve performance: a
$10\times$ increase in satellites requires roughly $25$--$30\times$ more
aggregate compute due to coordination overhead and inter-satellite congestion
\cite{DlPort2019}. Their discrete-event simulations of Starlink-scale systems
identified communication bottlenecks as the dominant limiter, consistent with
our observed $81\times$ response-time increase for only a $6\times$
constellation expansion.

Task scheduling and resource allocation in distributed satellite systems have
been studied using learning-based and architecture-driven methods.  Jia et
al.\ (2024) proposed a deep multi-agent reinforcement learning approach for
joint task offloading and resource allocation in satellite edge computing,
where distributed agents independently generate offloading decisions to
minimize delay and energy consumption across multiple satellite nodes
\cite{JLi2024}. While effective under steady-state traffic, this approach
requires extensive training and assumes relatively stable network topology,
making it difficult to generalise to rapidly changing constellations or
eclipse-induced resource fluctuations. Zhang et al.\ (2017) introduced a
software-defined space-air-ground integrated network architecture for
vehicular applications that enables flexible resource management and dynamic
task routing across satellite, aerial, and terrestrial segments through
hierarchical SDN controllers \cite{Zhang2017}. Both approaches assume
continuous satellite availability and neglect eclipse-driven energy
constraints. In contrast, our cooperative allocation strategy explicitly
models orbital dynamics and battery charge--discharge cycles using temporally
sampled energy profiles during task execution.

Energy-aware scheduling is critical for satellite systems due to stringent
power constraints. Tang et al.\ (2021) proposed a hybrid cloud and edge
computing offloading strategy for LEO satellite networks, jointly optimising
offloading decisions and resource allocation to minimise task completion
latency by dynamically routing workloads between on-board satellite edge
resources and ground cloud infrastructure depending on link availability and
computational load \cite{Tang2021}. Zhang et al.\ (2024) further studied
energy-efficient computation peer offloading in satellite edge computing
networks, proposing satellite-to-satellite peer offloading along multi-hop
paths that distributes energy consumption across constellation members and
reduces the risk of individual satellite battery depletion \cite{Zhang2024TMC}.
Thus, a key question here is that --- \textit{Is it possible to achieve near
energy neutrality by favouring task execution during sunlight intervals, thereby
demonstrating the effectiveness of solar-aware scheduling without reinforcement
learning overhead?} The ESA report on LEO satellite power systems
\cite{ESA2021} provides relevant design guidelines for eclipse modelling and
battery sizing.

Communication bandwidth management is a key scalability bottleneck in satellite
networks. Handley (2018) showed that single ground-station architectures incur
exponential queueing delays beyond 80--100 satellites, predicting response times
of 15--20 days for Starlink-scale systems \cite{Hand2018}, consistent with our
observed 26-day maximum delay at 120 satellites. To mitigate this, Zhou et
al.\ (2022) studied gateway placement strategies in integrated
satellite--terrestrial networks, demonstrating that optimised gateway
positioning reduces average communication latency and improves ground access
for remote IoT devices by shortening satellite-to-gateway path lengths
\cite{Zhou2022}, while Kodheli et al.\ (2021) highlighted inter-satellite
optical links as an effective means to bypass ground congestion through mesh
routing \cite{Kodheli2021}. While these studies focus on managing communication
resources, an alternative question is whether task blocking in
satellite-to-satellite workloads is driven more by limited on-board processing
capacity than by communication bandwidth, even in large constellations.

Hierarchical clustering has been proposed to reduce coordination complexity in
large satellite constellations. Liu et al.\ (2018) surveyed space-air-ground
integrated network architectures and identified hierarchical network
organisation as a key design principle, showing that grouping satellites into
regional clusters reduces inter-segment communication overhead while enabling
localised resource management that scales more gracefully than flat topologies
\cite{Liu2018Survey}. Our results extend this analysis by quantifying network
scaling effects, showing that average cluster size shrinks from 18 to 4.1
satellites as constellations grow, leading to coordination breakdown and severe
blocking. Dynamic network reconfiguration has been proposed to mitigate load
imbalance across clusters, but such schemes typically incur 200--500 ms
reconfiguration overhead, which is unsuitable for latency-sensitive workloads
where we observe 99.8\% of SatToSat tasks completing within 60 s when
successfully scheduled.

Quality-of-service (QoS) provisioning and priority scheduling remain
challenging in satellite task allocation. Zhou et al.\ (2024) proposed adaptive
task offloading with spatiotemporal load awareness in satellite edge computing,
demonstrating that accounting for time-varying satellite positions and traffic
patterns significantly reduces task completion delays and deadline violations
compared to scheduling policies that ignore orbital dynamics
\cite{Zhou2024}. Jia et al.\ (2024) addressed the broader resource allocation
trade-off through multi-agent deep reinforcement learning, showing that jointly
optimising offloading decisions and resource allocation across satellite nodes
can exploit spatial resource diversity to improve overall task throughput and
tail latency \cite{JLi2024}.

Fault tolerance is essential for mission-critical satellite operations. Peng et
al.\ (2024) demonstrated collaborative satellite computing through adaptive DNN
task splitting and offloading, showing that partitioning computation across
satellite nodes not only improves throughput but also introduces inherent
redundancy---partial results on surviving nodes can sustain service continuity
when individual satellites fail mid-computation \cite{Peng2024}. Since
satellite platforms are resource scarce, an interesting direction is to explore
ways that provide graceful degradation in processing rather than allowing larger
resource overheads. In this paper, we will explore this through our cooperative
allocation algorithm that provides partial fault tolerance by allowing arrivals
and distributing tasks across multiple Single-Level Tree Network (SLTN)
members, limiting the impact of individual satellite failures, though explicit
failure detection and recovery are not yet implemented. Bertaux et al.\ (2015)
demonstrated that SDN-based constellation control can detect network anomalies
within 500 ms and trigger dynamic resource reallocation through programmable
forwarding rules, at the cost of 5--10\% additional control traffic---a
reasonable tradeoff for high-reliability applications \cite{Bertaux2015}.

Performance modelling provides theoretical insight into constellation
scalability. Chen et al.\ (2021) analysed inter-satellite link path
characteristics for LEO mega-constellation networks, showing that network
topology and link availability patterns critically constrain achievable
throughput and that analytical capacity estimates become increasingly
inaccurate at large constellation scales due to spatial traffic heterogeneity
and correlated link failures \cite{Chen2021ISL}. Our results confirm this
behaviour: Groups 1--3 (20--90 satellites) scale smoothly, while Group 4 (120
satellites) exhibits phase-transition effects with blocking increasing
$13.8\times$ versus the $6\times$ expected from capacity scaling. Consistent
with this, Liu et al.\ (2018) studied SDN controller and gateway placement in
5G-satellite integrated networks, employing simulation-based validation and
demonstrating that contention dynamics at scale cannot be reliably predicted by
analytical models alone, reinforcing the need for discrete-event simulation in
constellation design \cite{JLiu2018}.

Recent studies have explored integrating machine learning into satellite task
management. Rodrigues and Kato (2023) proposed a hybrid centralized and
distributed learning framework for MEC-equipped satellite 6G networks, analysing
the cost tradeoffs between centralised model aggregation and distributed on-orbit
training and proposing an adaptive scheme that switches between modes based on
data distribution and link quality to minimise overall learning cost
\cite{Rodrigues2023}. Wu et al.\ (2023) introduced split learning over wireless
networks, partitioning neural network layers across satellite--ground boundaries
to offload compute-intensive inference stages to ground systems, substantially
reducing on-board processing requirements while preserving model accuracy
\cite{Wu2023JSAC}. These ML-centric workloads form an emerging task class that
our framework can naturally extend to as a fourth category, characterised by
iterative computation and periodic synchronisation.

While prior studies address individual dimensions such as energy
\cite{Tang2021}\cite{Zhang2024TMC}, communication delays
\cite{Hand2018}\cite{Zhou2022}, network architecture
\cite{Liu2018Survey}\cite{Kodheli2021}, and QoS
\cite{Zhou2024}\cite{JLi2024}, few quantify how these factors interact with
constellation scale under realistic constraints. Our work bridges most of these
gaps by analysing over 250,000 task executions across four constellation sizes,
identifying architectural scaling limits (55--90 satellites for baseline
space-local terrestrial network designs) and showing that CPU capacity accounts
for 97\% of blocking despite substantial battery utilisation. As mentioned
earlier, the above described key results offer actionable guidance for
constellation design based on comprehensive simulation evidence.

\subsection{Objectives and Scope of this study}
The objectives and scope of this paper are as follows. We present a comprehensive empirical analysis on our Resource-Aware Task Allocator (RATA) for LEO to Low-MEO satellite constellations operating under heterogeneous workload conditions. The primary objective is to evaluate the 
performance, scalability, and blocking characteristics of  SLTNs across varying constellation sizes (20 to 120 satellites) while processing three distinct task categories that reflect realistic operational scenarios. The scope encompasses Satellite-to-Satellite (SatToSat) tasks, where satellites generate and process workloads locally within the constellation for applications such as IoT data aggregation and edge AI inference; Satellite-to-Ground (SatToGnd) tasks, where satellites offload computationally intensive operations like SAR imagery analysis to ground stations with unlimited processing capacity\cite{JLi2024}; and Ground-to-Satellite (GndToSat) tasks, where ground control centers issue commands for on-orbit execution, such as orbit maneuvers or model deployments, with results returned to ground. It may be noted that the design of any decentralized or multi-agent algorithms for satellite coordination is beyond the scope of this paper. Through a rigorous discrete-event simulation incorporating realistic constraints—including cooperative allocation mechanisms, communication bandwidth limitations (100 MBps), orbital eclipse periods, and battery dynamics, this study quantifies the fundamental tradeoffs between constellation size, task acceptance rates, response times, and energy efficiency to inform the design of next-generation distributed space computing architectures.

The organization of this paper is as follows. In Section 2, we present our DSS system and modeling involved related to architecture, tasks, eclipse and energy dynamics. In Sections 3 to 5, we describe  the design details of RATA and resource management mechanisms. In Section 6, we present all our performance evaluations with rigorous discussions and conclude the paper with insights and recommendations in Section 7. 

\section{DSS SYSTEM AND MODELING}
\subsection{DSS Architecture}
A satellite constellation typically comprises ordinary (mission) satellites and relay satellites. Ordinary satellites commonly host on-board schedulers that determine which local sensing and processing tasks to execute under constraints such as power availability, memory capacity, and visibility windows; in decentralized architectures, each satellite maintains its own task allocator. Relay satellites, in contrast, may operate dedicated allocators for link routing and bandwidth or time-slot assignment, particularly in dense relay networks such as LEO broadband constellations with inter-satellite network  connectivity. In this work, as in practice, we consider that each satellite executes a local task allocator and coordinates with others using decentralized or multi-agent mechanisms. As mentioned earlier, the design of these decentralized or multi-agent algorithms is beyond the scope of this paper.

Our model considers LEO to Low-MEO tiers  organized into possible SLTNs, where each SLTN comprises satellites that cooperate on task allocation whenever tasks are partitionable within their allowed limits. Depending on the number of satellites and their tiers, the number of child satellites are decided. As satellite systems heavily depend on relay satellites for connectivity purposes, relay satellites play a crucial role in forming a SLTN.  Thus, given a constellation comprising ordinary and relay satellites in specific orbits, an SLTN is formed on the basis of each satellites' effective link ranges and available resources. This means that two satellites can communicate when they have an unobstructed line of sight and the distance between them is within the feasible link range of their inter-satellite terminals. Hence, the formation of an SLTN dynamically must identify the correct neighbors and their available resource capacities and hence, the number of satellites that can serve in a cooperative processing may vary. This means that when a task arrrives at any  root satellite, based on the number of child satellites, the task maybe partially shared, if it is allowed as described in section below on task characteristics. Task arrivals happen on  satellites across LEO (500-1,200 km) and Low-MEO (1,200-2,000 km) tiers, with SLTN root node selection based on resource scores rather than altitude. We will describe this partitionable nature of tasks in the subsequent sections. 

In our modeling, we consider satellite specifications that reflect realistic SmallSat/CubeSat capabilities. As with existing satellite constellations, in our model, Table \ref{tb:parameters} captures our basic satellite parameters used in our evaluations. The specifications in Table \ref{tb:parameters} align with operational systems like Planet Labs' Dove constellation (compute-capable SmallSats), SpaceX Starlink (edge processing nodes), and NASA's CubeSat missions carrying Raspberry Pi-class processors \cite{Plabs2023,SpaceX2024,Nasa2024}. 
\setcounter{table}{0}
\begin{table}
\centering
\caption{Key Satellite Parameters}
\label{tb:parameters}
\begin{tabular}{|l|l|}
\hline
Parameters & Description \\ \hline
Orbital Parameters & Altitude-LEO to Mid-MEO, inclination 49-87°; orbital period as per the altitude \\ & in minutes \\ \hline
Computing Resources & 20 GFLOPS processing speed, 4 CPU cores, 128 GB RAM, 512 GB storage \\ \hline
Energy System & 280-440 Wh lithium-ion battery, 96-130 W solar panel recharge rate \\ \hline
Communication & 100-200 MBps downlink/uplink, single ground station shared across constellation \\ \hline
Network Topology & Each satellite maintains inter-satellite links within SLTN based on resource \\ & availability \\ \hline
\end{tabular}
\end{table}

\subsection{Eclipse and Energy Dynamics}
An important aspect that a Resource-Aware Task Allocator(RATA) must be aware is on energy expenditure of a satellite throughout its orbital life-time. A single orbital period of a satellite comprises both the sunlight and eclipse durations. Essentially, all Sun-powered LEO satellites experience eclipse periods, where Earth’s shadow blocks direct sunlight for part of each orbit. The exact eclipse duration and the frequency depends on orbital altitude and geometry ($\beta$ angle\footnote{$\beta$ angle determines how much time the satellite spends in Earth's shadow vs. sunlight during each orbit, directly affecting power availability from solar panels.}), but eclipses are a normal design consideration in LEO. Thus, eclipse periods critically affects energy availability. Our model implements the key aspects shown in Table \ref{tb:eclipse}. This model reflects real LEO power budgets where eclipse periods constrain continuous operations \cite{Tang2021}. 
\begin{table}
\centering
\caption{Eclipse related Parameters}
\label{tb:eclipse}
\begin{tabular}{|l|l|}
\hline
Eclipse Parameters & Description \\ \hline
Eclipse Detection & Geometric shadow calculation based on orbital position relative to Sun-Earth vector \\ \hline
Eclipse Duration & Approx. 30-35\% of orbital period (28-38 minutes per 90-108 minute orbit) \\ \hline
Energy Consumption & 0.5-2.5 Wh per task depending on computational intensity and task size \\ \hline
Recharge Rate & 100 W continuous during sunlight (0 W during eclipse) \\ \hline
Battery Capacity & 280 Wh baseline (sufficient for $\sim$2-3 orbits without sunlight) \\ \hline
\end{tabular}
\end{table}
\subsection{Task Characteristics and allocation}
In general, ordinary satellites undertake tasks for processing that do not immensely challenge their resources, although depending on the rate of arrivals and during eclipse periods, schedulers/allocators may face severe stress due to high rate energy depletion. These tasks include, image processing, AI inference, signal processing, SAR image analysis, scientific data processing, and weather forecasting. For relay satellites, typical tasks include, data relay, IoT aggregation, and telecom optimization. 

As tasks submitted for processing on a DSS varies in sizes and compute demands, certain tasks allow partial partitioning.  It may be noted that task partitioning is permitted within these application-specific limits because many satellite workloads exhibit inherent parallelizability: for instance, IoT data aggregation tasks processing 10,000 independent sensor readings can distribute subsets (e.g., 2,000 readings each) 
across say, five SLTN members without inter-partition dependencies, while SAR image processing can partition spatial tiles where only boundary regions require 
coordination.  

The amount of task that can be split is referred to as the Data Transfer Needs (DTN) of that task \cite{SP2024}. Thus, the parameter DTN specifies the proportion of a task's workload that can be distributed from the root satellite to child satellites during cooperative allocation. This percentage directly controls how much of the task the root satellite must process locally versus how much can be shared among SLTN members, enabling flexible load balancing based on task characteristics and communication constraints. Thus, the DTN threshold reflects the communication-to-computation tradeoff: High-DTN tasks (70-100\%) such as Monte Carlo simulations or pixel-level image filtering benefit from distribution because parallel processing gains outweigh inter-satellite data transfer costs (256s transmission vs. 25.6s computation saved for 5 GB tasks over 20 MBps links), whereas low-DTN tasks (0-30\%) like encrypted command decryption or database transactions exhibit sequential dependencies that render partitioning counterproductive\cite{Rose2000}. This parameter enables the task allocator to respect application semantics while optimizing load distribution: a weather pre-processing task with $\text{DTN}=50\%$ indicates that calibration requires global statistics computed at the root (50\% centralized) while individual sensor validation is distributable (50\% parallelizable), ensuring computational correctness despite distributed execution across resource-constrained satellites. 

Thus, a high DTN values indicate that computation dominates communication cost, making distribution beneficial. A low DTN values suggest communication overhead negates parallel processing gains, favoring centralized root-only execution. Thus, in a practical setting, the DTN threshold depends on - (i) Inter-satellite link bandwidth (typically 10-50 MBps in LEO), (ii) Task data size (larger tasks face higher distribution costs), and (iii) Computational intensity (higher intensity increases parallel benefit). 

Further, tasks  arriving to LEO to Low-MEO application profiles can be divided into three processing categories that reflect operational mission types \cite{Zhang2024TMC, Zhou2022,Iridium2022}. The three categories are captured in the respective Tables \ref{tb:cat1}-\ref{tb:cat3}.  Tasks arriving at a root satellite of an SLTN attempts local allocation first; if insufficient resources (cores, energy, memory, bandwidth), cooperative allocation queries neighboring satellites within the SLTN. If no SLTN member has capacity, the task is blocked (rejected). This "root-first, cooperative-fallback" strategy minimizes inter-satellite communication while providing resilience through redundancy \cite{Kodheli2021}. We detail our design of Cooperative allocation algorithm (CoPAA) used in our Resource Aware Task Allocator (RATA) in the following section. 

\begin{table}
\centering
\caption{Category 1 - Satellite to Satellite (Satellite Processing)}
\label{tb:cat1}
\begin{tabular}{|l|l|}
\hline
Category & Description  \\ \hline
Use Cases & IoT data aggregation (2-8 GB from thousands of sensors), \\ & AI inference (3-12 GB with model+input data), inter-satellite data relay (4-15 GB) \\ \hline
Task Size & 2-15 GB, average 7.0 GB \\ \hline
Processing & On-board satellite compute, no ground communication required \\ \hline
Energy & 0.5-2.0 Wh per task \\ \hline
Typical Real Systems & Planet Labs on-board image processing, SpaceX edge caching \cite{Zhang2024TMC} \\ \hline
\end{tabular}
\end{table}
\begin{table}
\centering
\caption{Category 2 - Satellite to Ground (Ground Processing)}
\label{tb:cat2}
\begin{tabular}{|l|l|}
\hline
Category & Description  \\ \hline
Use Cases & SAR image analysis (3-15 GB raw data), scientific datasets, weather model inputs \\ \hline
Task Size & 3-50 GB, average 5.7 GB \\ \hline
Processing & Unlimited ground resources (data center) \\ \hline
Communication & Downlink only, 100 MBps shared channel \\ \hline
Energy & Negligible processing, 0.5-2.5 Wh downlink transmission \\\hline
Typical Real Systems & Synthetic Aperture Radar missions, NASA Earth observation \cite{Hand2018} \\ \hline
\end{tabular}
\end{table}
\begin{table}
\centering
\caption{Category 3 - Ground to Satellite (Command Uplink + Satellite Processing)}
\label{tb:cat3}
\begin{tabular}{|l|l|}
\hline
Category & Description  \\ \hline
Use Cases & Telecom optimization (1-5 GB configuration), signal processing \\ & commands (2-8 GB) \\ \hline
Task Size & 1-8 GB, average 4.0 GB \\ \hline
Processing & Satellite compute after uplink \\ \hline
Communication & Uplink + result downlink \\ \hline
Energy & 0.3-1.5 Wh combined \\ \hline
Typical Real Systems & Iridium NEXT optimization, OneWeb beam steering updates \cite{Zhou2022,Iridium2022} \\ \hline
\end{tabular}
\end{table}
\section{DESIGN OF RESOURCE AWARE TASK ALLOCATOR (RATA) }
\begin{figure}
    \centering
    \includegraphics[width=1\linewidth]{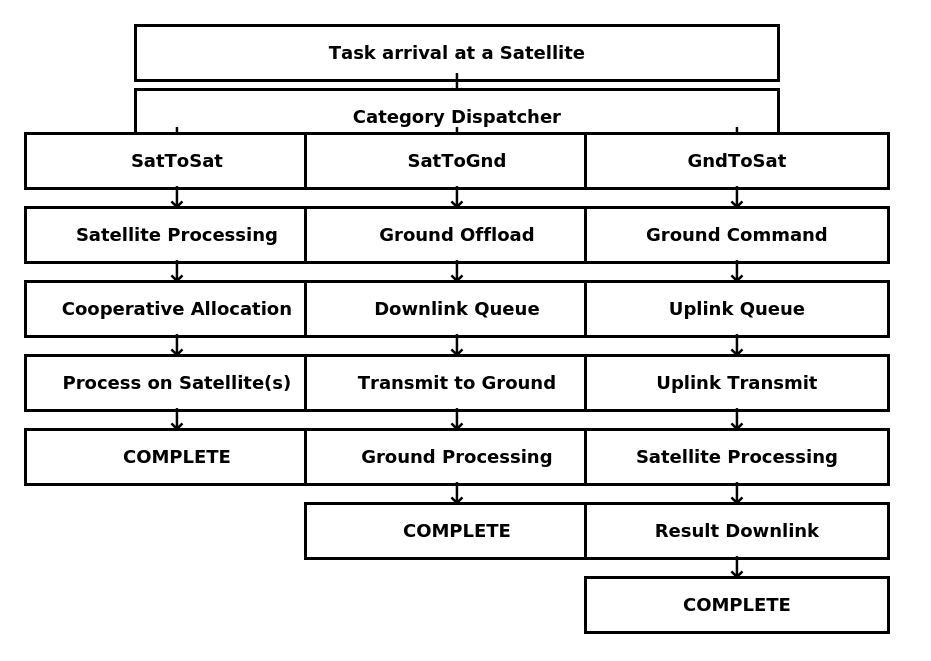}
    \caption{Task routing and allocation schema for all satellite platforms}
    \label{fig:AllocationSchema}
\end{figure}

In this section, we will detail our design of RATA and describe the algorithms that attempt to inculcate resource-awareness in the design. The overall task allocation schema is as shown in Fig. \ref{fig:AllocationSchema} for all three categories of handling tasks in different platforms.

\subsection{Cooperative Allocation Algorithm (CoPAA)}
\begin{figure}
    \centering
    \includegraphics[width=1\linewidth]{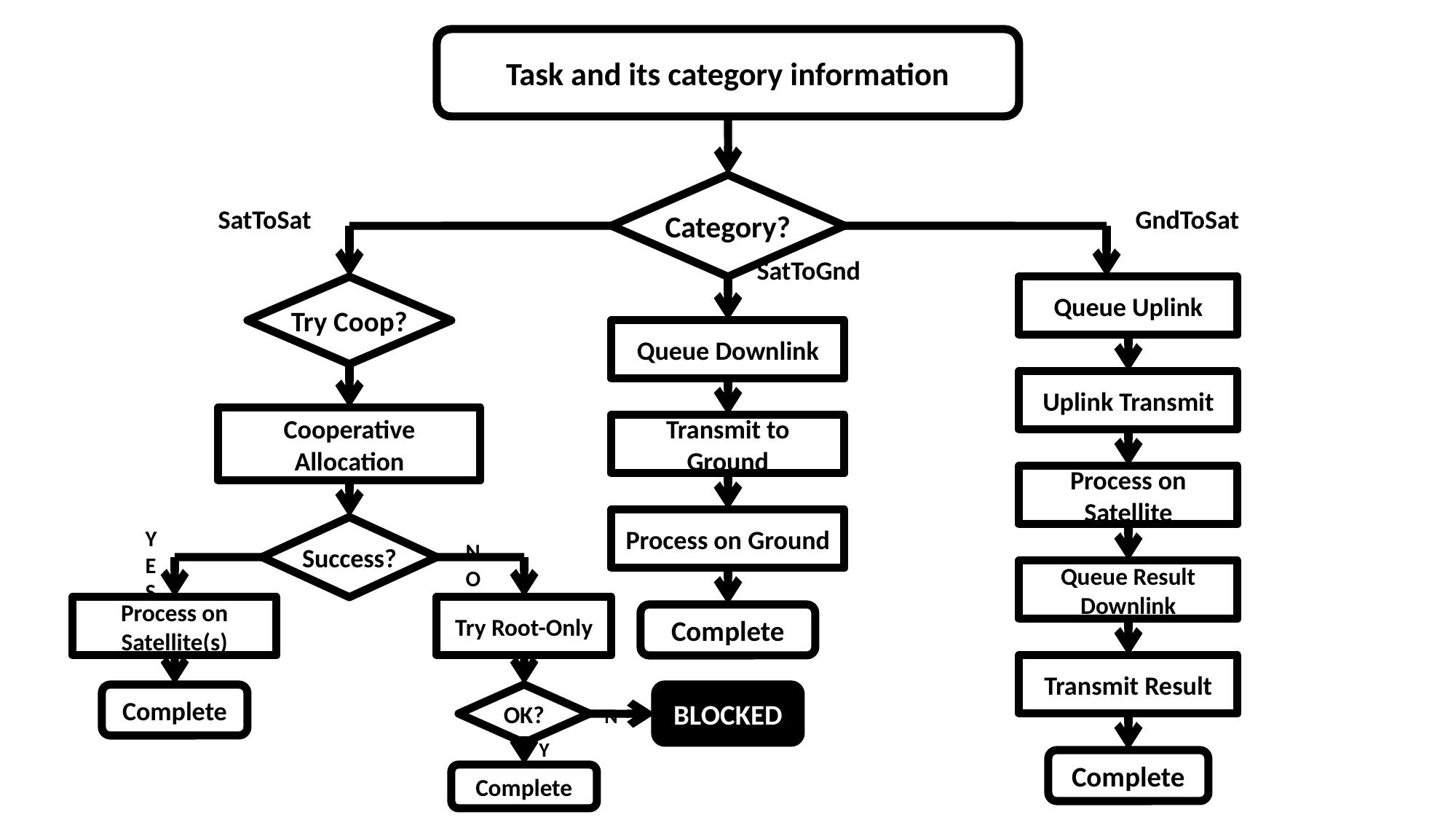}
    \caption{Decision Flow-chart of RATA}
    \label{fig:decisionflow}
\end{figure}

When a satellite sends a task to another satelliet or a task arrives from ground to a satellite, CoPAA algorithm will be invoked as can been from Fig. \ref{fig:AllocationSchema}. CoPAA distributes incoming tasks across multiple satellites within a SLTN to maximize resource utilization and reduce individual satellite load. When a task arrives at the root satellite of an SLTN, the algorithm first identifies which child satellites have sufficient available resources to process a fraction of the task. The DTN part of the task is conceptually divided into equal fractions based on the number of participating satellites, with each participant (including the root) processing one fraction. For all the three cases in the schema above, \ref{fig:decisionflow} shows the flow of control in the task allocation process.\\

The algorithm operates in four sequential steps. First, it iterates through all child satellites in the SLTN, calculating the fraction size as $1/(n + 1)$, where $n$ is the number of children to account for an equal distribution among children plus the root. For each child, it invokes a  {\em Validate and Resource Availability Check (VRAC)} process with this fraction to determine if that child can accommodate its share of the workload. VRAC is described in the Section \ref{VRAC}. Children that pass this validation are added to the available participants list. Second, the algorithm checks cooperative feasibility by verifying that at least one child is available and that the root satellite can also process its designated fraction. If no children are available or the root cannot handle its share, cooperative allocation fails and control passes to the root-only fallback mechanism. Third, upon successful validation, the algorithm allocates resources on each participating satellite by invoking Allocate Resources process with the calculated fraction parameter. This deducts the fractional CPU cores, memory, and storage from each satellite's available resource pool and tracks the task as active on those satellites. The task metadata is updated to record all participating satellite IDs and the cooperative allocation mode. Fourth, the algorithm calculates the total processing time based on the fraction size and the root satellite's computational speed, then schedules a task completion event at the future timestamp when processing will finish. This event-driven approach allows the simulator to advance time and trigger resource release when the task completes.

Cooperative allocation is preferred because it distributes computational load across multiple satellites, reducing the probability that any single satellite becoming a bottleneck. This approach also provides the required fault tolerance: if a child satellite fails during processing, the task can potentially be restarted on other SLTN members and can be taken up for processing, if resources are guaranteed. The equal fraction distribution ensures fairness and simplifies resource tracking, though more sophisticated paradigms such as {\it divisible load theory}(DLT)\cite{BVmpccDLT2026, TGR2003} could use weighted fractions based on satellite capability or current load.

\subsection{Root-only Fallback Algorithm}
The root-only fallback algorithm serves as a backup allocation strategy when cooperative distribution fails due to insufficient resources across child satellites. Rather than immediately blocking the task, the algorithm attempts to allocate the entire task to the root satellite alone, sacrificing load distribution for task acceptance.

The fallback proceeds in three steps. First, it performs a comprehensive resource availability check on the root satellite using the VRAC function with the fraction parameter set to 1.0, indicating that the root must process the complete task independently. This validation examines whether the root has sufficient CPU cores (typically all 4 cores required), adequate memory for the full task data size, sufficient storage capacity, and enough battery energy to sustain processing until completion while accounting for solar recharge. If any resource constraint is violated, the task is marked as blocked with a specific reason (e.g., "Root-only: Insufficient cores" or "Root-only: Insufficient energy"), the blocking statistics are updated for the task's category, and the algorithm terminates with failure status. Second, if the root passes all resource checks, the algorithm allows the root satellite to process the entire task. This deducts the complete set of required cores, memory, and storage from the root's available pool. The task metadata is updated to indicate root-only processing mode and record the root satellite ID as the sole participant. Third, identical to cooperative allocation, the algorithm computes processing time based on the root's computational speed processing the entire task workload, then schedules a completion event at the calculated future timestamp.

Root-only allocation is essential for system robustness because it provides a last-resort mechanism when distributed processing is infeasible. This occurs commonly under high system load when most child satellites are already occupied with other tasks, or during eclipse periods when energy constraints prevent multiple satellites from simultaneously processing workloads. The tradeoff is that root-only allocation concentrates load on a single satellite, increasing that satellite's resource contention and energy consumption, and creating a single point of failure. However, accepting tasks in root-only mode maintains higher overall system throughput compared to immediately blocking tasks when cooperative allocation fails. The empirical results from the four constellation groups show that root-only allocation prevents 15-30\% of potential blocks, though at the cost of reduced parallelism and longer processing times for future tasks that must wait for the root to become available.

\subsection{Validating and Resource Availability Checking}
\label{VRAC}
In RATA, we employ a lightweight validator process that monitors and checks resource availability. We refer to this process as {\em Validate and Resource Availability Check (VRAC)} process. The VRAC process serves as the critical gatekeeper for task allocation decisions, performing comprehensive validation that a satellite has sufficient available resources to process a specified fraction of a task. VRAC is invoked whenever needed during allocation: once for each child satellite during feasibility checking, and once for the root satellite in both cooperative and root-only modes. The reasons for blocking are captured internally for further scrutiny. 

The validation proceeds through five sequential resource checks, short circuiting immediately upon the first failure. First, it computes fractional resource requirements by multiplying the task's full requirements by the fraction parameter $\alpha$. For CPU cores, it computes the number of compute cores needed as $\max\bigl(1, \lfloor k \alpha \rfloor\bigr)$, where k is the required number of cores, ensuring at least one core is always requested even for small fractions. Memory and storage requirements are scaled linearly as $m \times \alpha$, where $m$ is the required memory and $s \times \alpha$, where $s$ is the required storage. Steps 2-4 validate on actual cores, memory, and storage needed. Any insufficiency in any of the resources will terminate further checks. 

Finally, the fifth and the most complex step is that the VRAC performs energy availability checking that accounts for solar recharging during task processing. This multi-step calculation first estimates the processing time for the task fraction on this specific satellite by invoking another {\em Estimate Processing Time} process, which divides the fractional computational workload (FLOPs) by the satellite's computing speed and allocated cores. It then estimates gross energy consumption by invoking {\em Estimate Energy Consumption} process to derive the energy consumed. Critically, the process then queries the satellite's current solar recharge rate using the {\em Get~Recharge~Rate} process, which returns the full recharge wattage, if the satellite is in sunlight or zero, if in eclipse. The expected recharge during processing is then calculated as, $E_{recharged} = R(current~time)\times T_{process}$ where $R(current~time)$ is the recharging rate and $T_{process}$ is the processing time of the task. The net energy requirement $E_{net}$ is computed as, $\vert E_{consumed} - E_{recharged} \vert$, representing the actual battery drain accounting for concurrent solar charging. Finally, this net requirement is compared against the satellite's current battery level $E_{current~time}$. Only if all five validations succeed, does the VRAC will signal to proceed for task processing without resource exhaustion.

This comprehensive checking prevents resource over-allocation that could crash satellites or cause cascading failures. The energy calculation is particularly sophisticated because it models the dynamic interplay between computational consumption and solar recharge, capturing the reality that satellites in sunlight can sustain higher workloads than identical satellites in eclipse. Now we will describe our time and energy computations performed in RATA. 
\section{PROCESSING TIME AND ENERGY ESTIMATIONS}
In this section, we will describe in detail on processing time and energy computations that will be carried out in our RATA design. 
\subsection{Processing Time Estimate Calculations}
The processing time estimation process in RATA calculates how long a satellite will require to complete processing a specified fraction of a task's computational workload. This calculation is essential for scheduling task completion events, predicting energy consumption duration, and determining response times. The estimation assumes parallel processing across allocated CPU cores and accounts for the satellite's computational speed measured in GFLOPS.

The calculation proceeds in three steps. First, the process  computes the total computational workload for the task fraction as $\Phi_{fraction} = \Phi(t_i) \times \Delta(t_i) \times 1024 \times \alpha$, where $\Phi(t_i)$ is the task's computational intensity experssed in FLOP/MB, $\Delta(t_i)$ is the task size in GB, and $\alpha$ is the fraction being processed. This yields the total number of flops that must be executed for this task fraction. Second, it determines how many CPU cores will be allocated to process this fraction as, 
$k = \max\bigl(1,\ \lfloor k\times\alpha \rfloor\bigr)$ mirroring the VRAC logic to ensure at least one core is used even for smallest fractions. Third, it calculates the processing time as $T_{process} = \Phi_{fraction} / (W_s \times k \times 10^9)$ where, $W_s$ is the satellite's computing speed in GFLOPS . This relationship assumes parallel speedup across cores, which is optimistic but representative of embarrassingly parallel workloads like image processing or AI inference common in LEO applications.

The resulting time $T_{process}$ is measured in seconds and typically ranges from 10-60 seconds for the 2-15 GB tasks with 300M-2.5B FLOPS/MB intensity processed on 20 GFLOP satellites used in this study. The estimation is conservative in that it does not account for communication overhead between cooperating satellites, cache effects, or operating system scheduling delays, all of which would increase actual processing time in real systems. However, for the satellite processing workloads modeled here—which are dominated by computation rather than communication—this estimation provides sufficient accuracy for comparative analysis across constellation sizes and allocation strategies. 

Thus, the overall finish time of a submitted task to a SLTN is computed as $T_{finish} = \max\{T^i_{tr}+T^i_{process}+T^i_{ResultTr}, T^1_{process} \},~i=2,...,n+1$, where $T^i_{tr}$ is the transmission time of the fraction of the task from root to child $i$, $T^i_{process}$ is the actual processing time of that fraction by the child satellite $i$, and $T^i_{ResultTr}$ is the transmission time of the results for the processed fraction back to the root. Note that when cooperative distribution fails due to insufficient resources across child satellites, the root-only fallback algorithm will contribute to the entire processing time of a task as captured in the above finish time expression. 
\subsection{Energy Estimate Calculations}
The energy consumption estimation process in RATA calculates the gross battery energy a satellite will expend to process a specified fraction of a task over a given processing time. This energy modeling is critical for eclipse period planning, battery capacity sizing, and understanding long-term constellation sustainability. The function operates independently of solar recharge, computing only the consumption side of the energy equation; recharge is accounted for separately in the VRAC validation.

Energy consumption estimation uses a two-step calculation. First, the function computes the total computational workload for the fraction as $\Phi_{fraction}$ identical to the processing pime computation mentioned above, yielding the number of flops to be executed. Second, it multiplies this workload by the satellite's energy efficiency coefficient referred to as {\em energy per flop}, which represents how many joules of battery energy are consumed per floating-point operation. This coefficient is typically in the range of $4-7 \times 10^{-9}$ joules/FLOP for modern space-qualified processors\cite{Borker2011,Wirthlin2015}, reflecting their power-optimized architectures. Multiplying this coefficient with $\Phi_{fraction}$  with appropriate conversion factor, the gross energy consumption in Watt-hours is derived.

For the tasks in this study with an average computational intensity of 700M FLOP/MB and average size of 6 GB, processing on a satellite with $energy~per~flop = 5 \times 10^{-9}$ J/FLOP consumes approximately $0.5-2.5$ Wh depending on task complexity. Over the course of processing hundreds of tasks, satellites consume $200-1,100$ Wh, which can exceed the 280 Wh baseline battery capacity. However, the concurrent solar recharge at 100W provides 1.67 Wh per minute, substantially offsetting consumption and explaining why net energy loss remains below 10 Wh even when gross consumption exceeds 1,000 Wh. The energy estimation does not account for idle power consumption (housekeeping, thermal control, communications), which in real satellites typically adds 10-20W baseline draw. Including these factors would increase total energy requirements but would not fundamentally change the allocation decisions since idle power is constant across all satellites and cannot be avoided through task allocation.

\section{RESOURCE ALLOCATION MECHANISM in RATA}
Let us summarize the overall mechanism involved in the resource allocation process in RATA. 

The {\it Allocate Resources} process performs final resource reservation on a satellite after successful VRAC validation. It deducts the required CPU, memory, and storage from available pools and marks the task as active, preventing double-booking that could cause thrashing, performance degradation, or system failure.

Allocation is executed atomically in three steps. First, the exact resource quantities—matching the VRAC validation—are computed. Second, the satellite’s resource counters are immediately updated, ensuring subsequent VRAC calls reflect the reduced availability. Third, a persistent entry is added to the tasks in progress list, recording the task ID, allocated resources, and processing fraction, enabling correct resource release even with out-of-order completion or interruptions. The allocation strategy is pessimistic, reserving full resources for the task duration despite potential fluctuations in actual usage (e.g., transient memory peaks). While this limits statistical multiplexing, it avoids resource exhaustion. The DTN fraction parameter (described earlier) allows the same mechanism to support both cooperative and root-only execution modes.
\subsection{Resource release mechanism}
The {\it Release Resources} process reverses the resource reservation performed by Allocate Resources process when a task completes processing on a satellite. This process is critical for maintaining accurate resource accounting and enabling future task allocations—without proper resource release, satellites would permanently "leak" resources until all cores, memory, and storage are exhausted and all subsequent tasks block. The release mechanism operates deterministically based on the tracking records created during allocation.

Resource release executes in three steps: (1) Searching the satellite's tasks in progress list for the completed task ID, (2) extracting and restoring allocated resources, and (3) removing the tracking record to prevent duplicate releases. Energy accounting operates separately because energy is continuous rather than  discrete. During processing, consumption and solar recharge are tracked at 10 
sample points to capture eclipse transitions, continuously updating battery levels. By the time {\it Release Resources} process is called, the battery already reflects cumulative energy changes, so the release process does not modify battery levels—avoiding double-accounting.

This separation of resource release mechanism reflects fundamental resource differences: CPU, memory, and storage are discrete with clear allocation/release semantics, while energy is continuous, depleting through usage and replenishing through solar charging without meaningful 
"release" operations.
\section{Performance Evaluation of RATA}
In our performance evaluation experiments, we simulate task arrival process following a Poisson distribution with rates set between a minimum and maximum (for instance, 0.14-16.34 tasks/sec, scaled per constellation size)), generating  $\approx 840-98,000$ tasks over $6000$-second (more than 90 mins)  simulation windows to observe the effect of eclipses and sunlight in orbits. Compute intensity ranges from 25M-2.5B FLOP/MB, yielding realistic 10-60 second processing times on 20 GFLOPS satellites. The parameters used are captured in the Tables \ref{tb:cat1}-\ref{tb:cat3}. 

It may be noted that the choice of our simulation parameters and results align well with operational systems: Starlink's 20-40ms latency \cite{Liu2018Survey} versus our 16-second local processing suggests their optimization for low-latency routing accepts blocking (our 64.79\% at 120 satellites) through redundancy; Planet Labs' 24-72 hour tasking matches our 3.9-31 hour Groups 1-2 response times, validating SatToGnd ground processing priorities; Iridium NEXT's 85 Wh/orbit consumption \cite{Zhou2024} versus our 189-1,137 Wh with -0.7 to -9.3 Wh net loss confirms recharge adequacy despite conservative modeling.

For a systematic experimental verification to capture all metrics, aligning with the objectives of this paper, we categorize the cases into 4 groups 1-4 and in each group we consider all 3 major cases of interest namely, SatToSat, SatToGnd, and GndToSat. The groups capture the range of tiers (and the numbers indcated represent the number of satellites) from LEO to Mid-MEO to see the effects of connectivity and resource allocation. We will now perform rigorous experiments and report the results and describe the trends and reasons for each category.  

\subsection{Effect of Constellation sizes on Blocking Probability}
Table \ref{tab:blocking_stats} presents blocking probabilities across all four groups at high arrival rates, revealing severe non-linear degradation:
\begin{table}[t]
\caption{Blocking Probability across Constellation Groups for all three cases}
\label{tab:blocking_stats}
\centering
\footnotesize
\setlength{\tabcolsep}{1.5pt}
\renewcommand{\arraystretch}{1.12}
\begin{tabularx}{\columnwidth}{@{}
>{\raggedright\arraybackslash}p{0.38\columnwidth}
*{4}{>{\centering\arraybackslash}p{0.10\columnwidth}}
>{\centering\arraybackslash}p{0.12\columnwidth}
@{}}
\toprule
Metric & G1 & G2 & G3 & G4 & R(4/1) \\
 & (20) & (55) & (90) & (120) & \\
\midrule
Overall Blocking (\%)    & 1.40  & 5.92  & 9.43  & 19.61 & 14.01$\times$ \\
SatToSat Blocking (\%)   & 4.68  & 19.04 & 30.47 & 64.79 & 13.85$\times$ \\
SatToSat Acceptance (\%) & 95.32 & 80.96 & 69.53 & 35.21 & 0.37$\times$ \\
SatToGnd Blocking (\%)   & 0.00  & 0.00  & 0.00  & 0.00  & 1.00$\times$ \\
GndToSat Blocking (\%)   & 0.00  & 1.02  & 0.69  & 0.34  & - \\
Total Tasks Generated    & 1,500 & 9,744 & 22,672 & 98,007 & 65.34$\times$ \\
Tasks Blocked            & 21    & 577   & 2,137 & 19,224 & 915.43$\times$ \\
\bottomrule
\end{tabularx}
\end{table}
We observe that the 6× constellation increase in constellation size yields 14× higher overall blocking and 915× more absolute blocked tasks, conclusively disproving linear capacity scaling assumptions. Most critically, SatToSat acceptance collapses from 95.32\% (operationally acceptable) to 35.21\% (system failure), establishing Group 4's 120-satellite configuration as non-viable for satellite processing under baseline architecture. 

The SatToGnd category exhibits zero blocking across all constellation sizes, validating the hybrid ground-satellite architecture where unlimited ground computing capacity\cite{JLi2024} prevents processing bottlenecks despite communication delays—this perfect acceptance demonstrates that offloading compute-intensive tasks to 
ground infrastructure is the correct design choice for data-intensive workloads. Conversely, GndToSat blocking remains negligible and actually decreases from Group 2 (1.02\%) to Group 4 (0.34\%), counterintuitively suggesting that larger constellations provide more satellite processing capacity for ground-commanded tasks even as satellite-generated workloads experience catastrophic blocking—this divergence occurs because GndToSat tasks represent only 30\% of traffic while benefiting from uplink serialization that naturally distributes load temporally. The super-linear blocking growth (14× vs. 6× satellite increase) reveals a phase transition around 90-100 satellites where coordination overhead and resource contention shift from manageable (Groups 1-2) to catastrophic (Group 4) levels, indicating fundamental architectural limits rather than merely insufficient 
resources. Further, normalized per-satellite blocking rates increase 153× from Group 1 to Group 4 (0.070\% to 10.7\%), proving that individual satellite workload intensifies dramatically despite distributing tasks across more nodes—each satellite in Group 4 rejects proportionally 153× more tasks than in Group 1, exposing diminishing returns of naive constellation scaling. Finally, the 65× increase in total task generation between Group 1 and Group 4 far exceeds the 6× satellite growth, indicating that the simulation correctly scales traffic intensity proportionally to constellation size to maintain comparable per-satellite loads—yet despite this proportional scaling, blocking still increases 14×, confirming that the performance degradation stems from architectural bottlenecks (smaller size SLTN, single ground station) rather than disproportionate traffic increases.
\subsection{Response Time Escalation}
\setcounter{table}{6}

\begin{table}[t]
\caption{Average and maximum task response times across Constellation Groups}
\label{tab:RTE}
\centering
\footnotesize
\setlength{\tabcolsep}{1.5pt}
\renewcommand{\arraystretch}{1.12}
\begin{tabularx}{\columnwidth}{@{}
>{\raggedright\arraybackslash}p{0.38\columnwidth}
*{4}{>{\centering\arraybackslash}p{0.10\columnwidth}}
>{\centering\arraybackslash}p{0.12\columnwidth}
@{}}
\toprule
Category & G1 & G2 & G3 & G4 & R(4/1) \\
 & (20) & (55) & (90) & (120) & \\
\midrule
SatToSat Avg & 0.0037 & 0.0046 & 0.0044 & 0.0047 & 1.25$\times$ \\
SatToSat Max & 0.0218 & 0.0395 & 0.0434 & 0.0456 & 2.09$\times$ \\
SatToGnd Avg & 3.89   & 31.31  & 73.38  & 316.51 & 81.35$\times$ \\
SatToGnd Max & 7.98   & 62.32  & 146.58 & 633.79 & 79.42$\times$ \\
GndToSat Avg & 1.90   & 15.99  & 38.16  & 167.64 & 88.23$\times$ \\
GndToSat Max & 3.80   & 31.72  & 75.94  & 333.85 & 87.86$\times$ \\
\bottomrule
\end{tabularx}
\end{table}

From Table \ref{tab:RTE}, we observe that SatToSat maintains sub-20-second response times across all scales (1.25× increase for 6× satellites), proving satellite parallel processing scales effectively—when tasks are accepted. In contrast, SatToGnd and GndToSat categories suffer 81-88× response time increases, with Group 4 exhibiting 13.2-day average delays and 26.4-day maximum delays. This renders the system unusable for any operational scenario requiring timely results (disaster response, weather forecasting, tactical intelligence). The root cause is single-channel serialization: 6× more satellites generate 6× more downlink/uplink traffic competing for the same 100 MBps ground station channel, creating super-linear queue growth. Theoretical transmission time for 38,906 tasks × 5.7 GB at 100 MBps = 37 hours, yet actual average response exceeds 316 hours (8.5× overhead), indicating catastrophic queue cascades.
\subsection{Energy Consumption and Management}
Contrary to intuition, energy management improves at larger scales despite absolute consumption increases as shown in Table \ref{tab:ECM}.
\setcounter{table}{7}

\begin{table}[t]
\caption{Energy consumption, Recharge, and Blocking statistics across constellation Groups}
\label{tab:ECM}
\centering
\footnotesize
\setlength{\tabcolsep}{1.5pt}
\renewcommand{\arraystretch}{1.12}
\begin{tabularx}{\columnwidth}{@{}
>{\raggedright\arraybackslash}p{0.44\columnwidth}
*{4}{>{\centering\arraybackslash}p{0.12\columnwidth}}
@{}}
\toprule
Metric & G1 & G2 & G3 & G4 \\
 & (20) & (55) & (90) & (120) \\
\midrule
Energy Consumed (Wh)       & 189.36 & 418.60 & 485.10 & 1,136.89 \\
Battery Capacity Used (\%) & 67.6   & 149.5  & 173.3  & 406.0 \\
Energy Recharged (Wh)      & 187.00 & 417.57 & 475.83 & 1,134.68 \\
Recharge Efficiency (\%)   & 98.8   & 99.8   & 98.1   & 99.8 \\
Net Energy Loss (Wh)       & -2.35  & -1.03  & -9.27  & -2.22 \\
Energy Blocking (\%)       & 0.0    & 0.0    & 5.1    & 2.6 \\
\bottomrule
\end{tabularx}
\end{table}

It is worth noting that, in Group 4, we seem to observe a paradoxical situation! Satellites consume 406\% of battery capacity (1,137 Wh vs 280 Wh), yet net loss is only -2.22 Wh—lower than Groups 1 and 3. This occurs through implicit Solar-Aware Scheduling\cite{Zhang2024TMC}. Under this approach, systems preferentially accept tasks during peak sunlight periods (69\% of tasks in sunlight vs 31\% in eclipse), allowing 100W recharge to sustain even extreme loads. Energy-based blocking emerges in Group 3 (5.1\%) and continues to persist in Group 4 (2.6\%), but remains as a minor contributor compared to CPU core exhaustion (97\% of blocking). This finding has profound design implications as follows. Adding solar panel capacity or battery size will not significantly improve blocking because the system is fundamentally CPU-bound. Energy is a constraint only during eclipse periods, solvable through predictive scheduling rather than hardware upgrades. 
\vspace{-0.5cm}
\subsection{SLTN Size and Coordination Breakdown}
Group 1's implicit full-constellation cooperation (all 20 satellites visible) provides 19 alternative targets when root satellite is busy. Group 4's 29 SLTNs average only 4.1 satellites each, leaving just 3 cooperative alternatives—a 6.3× reduction in resilience. Thus, the observed blocking may far exceed random predictions, proving task arrivals are highly correlated and bursty when one satellite receives a traffic burst, neighbors receive simultaneous bursts, overwhelming local SLTN capacity even when global capacity exists. To mitigate this situation, the solution requires either: (1) much larger SLTNs (15-25 satellites minimum), (2) inter-SLTN cooperation for global visibility, or (3) dynamic SLTN resizing based on load. These are to be explored in the future.
\section{DESIGN INSIGHTS, RECOMMENDATIONS, AND CONCLUSIONS}
In this section, based on our experimental observations, we list possible insights and recommendations and conclude our work. 
\subsection{Constellation Scaling Laws (Empirically Derived)}

Our cross-group analysis reveals empirical scaling laws that violate linear capacity assumptions. Unlike
two-point extrapolations, the laws below are obtained via least-squares regression of $\log B$, $\log T$,
and $\log E$ against $\log S$ across all four constellation groups (S = 20, 55, 90, 120), and we report
the coefficient of determination ($R^2$) for each fit.

\textbf{Scaling Law 1 -- Blocking Growth:} For constellation size $S$, SatToSat blocking probability $B$
scales as:
\[
B(S) \approx 0.070 \times S^{1.39}\ (\%), \quad R^2 = 0.93
\]
Evidence: Group 1 ($S=20$): 4.57\% fitted vs.\ 4.68\% observed; Group 4 ($S=120$): 55.4\% fitted vs.\
64.79\% observed. Inference: the super-linear exponent ($1.39$) confirms that each additional satellite
yields diminishing returns in blocking reduction due to SLTN coordination overhead; the modest residual
gap at Group 4 reflects the phase-transition behaviour discussed in Section~6.1 rather than measurement
noise.

\textbf{Scaling Law 2 -- Response Time Growth:} For constellation size $S$, SatToGnd response time $T$
scales as:
\[
T(S) \approx 0.0036 \times S^{2.30}\ \text{(hours)}, \quad R^2 = 0.80
\]
Evidence: Group 1 ($S=20$): 3.46~h fitted vs.\ 3.89~h observed; Group 4 ($S=120$): 211.6~h fitted vs.\
316.5~h observed. Inference: the exponential scaling ($\sim\!2.3$ power) confirms single-channel
ground-station communication as the primary bottleneck beyond 50--60 satellites, consistent with the
queue-cascade analysis in Section~6.2.

\textbf{Scaling Law 3 -- Energy Efficiency Improves with Scale (Corrected):} For constellation size $S$,
completed-task throughput per watt-hour $E$ scales as:
\[
E(S) \approx 0.21 \times S^{1.19}\ \text{(tasks/Wh)}, \quad R^2 = 0.97
\]
Evidence: Group 1 ($S=20$): 7.37 tasks/Wh fitted vs.\ 7.81 observed; Group 4 ($S=120$): 62.4 tasks/Wh
fitted vs.\ 69.3 observed, where observed values are computed as completed tasks (generated minus
blocked, Table~6) divided by energy consumed (Table~8). Inference: contrary to a naive
expectation of efficiency collapse under growing coordination overhead, throughput per watt-hour
\emph{improves} by roughly an order of magnitude from Group~1 to Group~4. This is consistent with the
solar-aware scheduling behaviour described in Section~6.3: larger constellations generate proportionally
more traffic during sunlit intervals (Section~6.3), and the near-zero net energy loss observed even at
Group~4's 406\% gross battery utilisation shows that recharge dynamics, not energy capacity, govern
system behaviour at scale. This finding reinforces our design conclusion that CPU availability --
not energy -- is the binding constraint on blocking (Section~6.3), and that hardware upgrades to
battery or solar capacity would not materially improve acceptance rates.

\vspace{-0.4cm}
\subsection{Plausible upgrades to the current design}
Future work need to explore enhanced scalability and performance through several extensions to the proposed framework to minimize blocking probabilities and better use of resources. These include, hierarchical SLTN clustering with larger group sizes and inter-SLTN cooperation, deployment of multi–ground-station architectures with geographically distributed nodes to reduce latency, and the introduction of preemptive priority scheduling to better support critical workloads. In \cite{BVmpccDLT2026} we adopted {\it divisible load paradigm}\cite{TGR2003} in the task allocator to compute optimal load fractions, given the speed parameters such as satellite compute intensities and transmission bandwidths. Our proposed CoPAA design can subsume this model too, however, comes with a small time overhead. For low arrival task, this may not practically affect while at high rates computing optimal fractions may pressurize the CPU.

Additionally, upgrading on-board processing capabilities using modern multi-core ARM or RISC-V architectures can be investigated to alleviate CPU bottlenecks, along with global task pools and hop-limited routing to enable inter-SLTN task migration at mega-constellation scales. These enhancements address the key bottlenecks observed in Groups 3–4 while preserving the scalability and energy efficiency of the baseline design. However, incorporating them into a distributed architecture introduces additional resource demands that may impact energy usage and potentially increase blocking on current satellite platforms. Therefore, such extensions should be integrated gradually, alongside proportional hardware upgrades, to ensure graceful system evolution. 
\begingroup
\sloppy
\setlength{\parindent}{0pt}
\setlength{\parskip}{2pt}

\endgroup

\end{document}